\documentclass[11pt,floatfix,aps,nofootinbib]{revtex4} 
\usepackage{amssymb,amsfonts}

\usepackage{epsfig}

\usepackage{color}

\usepackage{subfigure}
\usepackage{graphicx}
\usepackage{hyperref}
\usepackage{cancel}

\def\Tr{{\rm Tr}}

\def\beq{\begin{equation}}
\def\eeq{\end{equation}}

\def\bea{\arraycolsep .1em \begin{eqnarray}}
\def\eea{\end{eqnarray}}
\def\Tr{{\rm Tr}}

\def\eq#1{(\ref{#1})}

\def\s0#1#2{\mbox{\small{$ \frac{#1}{#2} $}}}
\def\0#1#2{\frac{#1}{#2}}

\def\grgl{\:\hbox to -0.2pt{\lower2.5pt\hbox{$\sim$}\hss}{\raise3pt\hbox{$>$}}\:}
\def\klgl{\:\hbox to -0.2pt{\lower2.5pt\hbox{$\sim$}\hss}{\raise3pt\hbox{$<$}}\:}

\begin{document}
\title{Black hole thermodynamics under the microscope}

\author{Kevin Falls}
\author{Daniel F. Litim}
\affiliation{Department of Physics and Astronomy, University of Sussex,
Brighton BN1 9QH, U.K.}
\date{\today}
\begin{abstract}
A coarse-grained version of the effective action is used to study the thermodynamics of black holes, interpolating from largest to smallest masses.
The physical parameter of the black hole are linked to the running couplings by thermodynamics, and
the  corresponding
equation of state includes quantum corrections for temperature, specific heat, and entropy.
If quantum gravity becomes asymptotically safe, the state function predicts 
conformal scaling in the limit of small horizon area, and bounds on black hole mass and temperature.
A metric-based derivation
for the equation of state, and
quantum corrections to the thermodynamical, statistical, and phenomenological definition of entropy are also given. 
Further implications and limitations of our study are discussed.
\end{abstract}
\maketitle
\tableofcontents
\section{Introduction}

The  discovery of black hole thermodynamics nearly half a century ago  points towards a deep connection between the laws of classical general relativity, thermodynamics and quantum mechanics due to the presence of causal horizons. Under the assumption that Einstein's equation holds true, it was noted  that  black hole solutions obey a set of laws analogous to those of thermodynamics \cite{Bardeen:1973gs} with the surface gravity $\kappa$ and the area of  the black hole horizon $A$ playing the r$\hat{\rm o}$les of temperature $T$ and entropy $S$, respectively \cite{Bekenstein:1973ur}. Including quantum mechanics, the quantitative link between area, entropy, Newton's coupling $G_N$ and Planck's constant $\hbar$ was found to be
\begin{equation}\label{entropy}
S=\frac{A}{4\,G_N\,\hbar}
\end{equation}
based on the thermal radiation emitted by quantum fields on black hole space-times \cite{Hawking:1974sw}. 
 Even more intriguingly, it was  also realized that the thermodynamic structure of gravity constitutes an {equivalence}, meaning that the line of reasoning can be reverted: provided that the fundamental thermodynamical relation $\delta Q=T\,\delta S$ between heat, temperature and entropy holds true for every local Rindler causal horizon, Einstein's equation follows as the corresponding equation of state  \cite{Jacobson:1995ab}. 
Hence, the semi-classical picture would seem to suggest that there exists an underlying micro-structure of space-time analogous to how  the thermodynamics of a gas follows from the ``coarse-graining" of  atoms or molecules. This brings up intriguing questions including whether metric gravity should be quantized at all, or rather thought of as an emergent macroscopic phenomenon 
\cite{Barcelo:2005fc,Padmanabhan:2009vy, Verlinde:2010hp,Percacci:2010af}.  

The Hawking effect also seems to introduce a new level of uncertainty into the laws of physics \cite{Hawking:1976ra}.
In particular a thermal bath of particles, as seen by distant observers, contains little information of the matter that initially collapsed to form the black hole and if left alone the black hole will eventually evaporate completely.
This process suggests that  pure states can evolve into mixed states, which is in conflict with the basic principle of a unitary time evolution in quantum mechanics. 
It would thus be important to understand  the deeper origin of the Bekenstein-Hawking entropy \eq{entropy}  in terms of microscopic degrees of freedom, and how fluctuations will affect the thermodynamical structure of gravity.

The notion of ``coarse-graining", as put forward by K.~Wilson \cite{Wilson:1973jj,Wilson:1975ly} and L.~Kadanoff \cite{Kadanoff:1966wm} nearly half a century ago, has become a key concept  in the understanding of quantum field theory and statistical physics. 
On the level of the path integral the idea
corresponds to the successive integrating-out of momentum modes. A virtue of these methods is that they permit a continuous interpolation between the microscopic and macroscopic degrees of freedom, controlled by exact functional renormalisation group equations for the effective action \cite{Berges:2000ew,Litim:1998nf}. 
 For gravity, the idea of a coarse-grained path integral has received increasing attention in the past decade \cite{Floreanini:1993na,Reuter:1996cp,Litim:2006dx,Niedermaier:2006ns,Niedermaier:2006wt,Percacci:2007sz,Litim:2008tt,Litim:2011cp,Reuter:2012id}, largely motivated by the quest for a gravitational fixed point within  S.~Weinberg's asymptotic safety conjecture  \cite{Weinberg:1980gg}. Applications to black holes cover RG-improved versions of Schwarzschild space-times in four dimensions \cite{Bonanno:2000ep} and their dynamics \cite{Bonanno:2006eu}, the thermodynamics of higher-dimensional black holes \cite{Falls:2010he} and their phenomenology \cite{Burschil:2010ys}, rotating black holes in four \cite{Reuter:2011uq} and higher dimensions \cite{Litim:2013gga},  black holes in higher-derivative gravity \cite{Cai:2010zh}, and RG flows from boundary terms \cite{Becker:2012js}.

In this paper we follow a novel route to explore the idea that a continuous coarse-graining of metric degrees of freedom could give rise to the thermodynamics as encoded in the horizon structure of black hole solutions. The questions we wish to address  with this is whether there exists a  version of black hole thermodynamics applicable for small, possibly Planck-size, black holes, and if so, what can be learnt from the corresponding equation of state? Following  closely the original derivation of black hole thermodynamics, our primary new addition is to replace the underlying action by a scale-dependent, coarse-grained effective action.  Thermodynamics then provides a link between the scale-dependence of couplings and the horizon area of black holes, leading to quantum corrections for black hole state functions.  This allows for a continuous interpolation between macroscopic black holes, where semi-classical results such as \eq{entropy}  serve as a reference point, and microscopic ones, which are informed by quantum corrections through the RG evolution of couplings.  Our results are obtained in a metric-independent fashion and apply for non-rotating, rotating, or charged black holes alike. We also show that a metric, which carries the same equation of state, can be obtained by means of an RG improvement of semi-classical space-times,  thereby closing a gap in the literature. 
In addition, we study the implications for black holes within asymptotically safe quantum gravity and derive quantum corrections to temperature and specific heat. Quantum corrections to the entropy are equally computed including those for a thermodynamical, statistical, and phenomenological definition of entropy, and contrasted with results from other approaches to quantum gravity. We also find that the conformal scaling of asymptotically safe gravity is encoded in the state function for small horizon areas.

The remaining part of the paper is organised as follows.  We briefly recall the basics of  black hole thermodynamics and the functional renormalisation group for gravity, and introduce notation and conventions  (Sec.~\ref{Preliminaries}). This is then followed by the construction of an RG improved version of black thermodynamics at the example of Kerr-Newman type black holes within Einstein-Maxwell gravity 
(Sec.~\ref{secII}).
We then specify to the case where gravity becomes weaker at high energies as predicted  by the asymptotic safety conjecture, and discuss RG corrections to temperature, specific heat, entropy,  and the occurrence of conformal scaling (Sec.~\ref{secIII}).
We also show that our set-up can be realized in terms of explicit RG improved black hole metrics, and discuss our results for the entropy and highlight similarities and differences with earlier studies (Sec.~\ref{secIV}). We end in Sec.~\ref{secV} with a brief discussion of results and some conclusions.

\section{Preliminaries} \label{Preliminaries}
In this section we provide some prerequisites for our study, including a 
brief overview of black hole thermodynamics and Wilson's 
renormalisation group. We also introduce some notation.

\subsection{Black hole thermodynamics}
In four dimensions stationary black hole solutions to the coupled  
Einstein-Maxwell equations are parameterised by their mass $M$, 
angular momentum $J$ and charge $q$ \cite{Carter:1971zc, Robinson:1975bv}, 
a result known as black hole uniqueness. This set of most general 
 black hole solutions for  long-ranged forces, the Kerr-Newman black holes, are expected to be the end points of gravitational collapse \cite{Wald:1971iw}. 
 It follows from black hole uniqueness that the area of the horizon may be 
 considered as a function $A=A(M,J,q)$.

 If an infinitesimal amount of matter crosses the horizon the area $A$ 
 of the horizon will vary according to \cite{Bardeen:1973gs} 
\beq \label{firstlaw}
   \frac{\kappa}{8 \pi G_N} \delta A = \delta M  - \Omega\, \delta J -  \Phi\, e\, \delta q
\eeq
 where $\kappa$, $\Omega$ and $\Phi$ are the surface gravity, 
angular velocity and electric potential evaluated at the horizon. Here $G_N$ 
and $e$ denote Newton's constant and elementary
 electric charge, respectively (we work in units $c=1$). In this paper we make
 the split $q \to e q$ where $q$ is the quantity of charge and $e$ is the coupling.
  The equation \eq{firstlaw} has the form of the first law of thermodynamics 
  $\delta U = \delta Q + \mu_i \delta N_i$ for which the internal energy $U$ is 
  associated to $M$, the heat crossing the horizon  $\delta Q$ is identified 
with $\frac{\kappa}{8 \pi G_N} \delta A$ and the conserved quantities $N_i$ 
and the associated chemical potentials $\mu_i$ with $\{J,q\}$  and $\{\Omega, \Phi\}$, 
respectively. As in conventional thermodynamics one can think of the black 
hole area $A$ as a ``state function" $A(M,J,q)$,  defining a set of states parameterised 
by $M$, $J$ and $q$. Then by taking appropriate derivatives in line with the first 
law \eq{firstlaw}, one can obtain the intensive quantities $\kappa$, $\Omega$ 
and $\Phi$. For an equilibrium thermodynamical  process at temperature $T$, the 
heat transfer $\delta Q$ due to the coarse-grained microscopic degrees of freedom 
is related to the change in entropy $\delta S$ by
\beq \label{Clausius}
\frac{\delta Q}{T} = \delta S \, .
\eeq 
Additionally the second law of thermodynamics states that the entropy of an isolated 
system can never decrease $\delta S \geq 0$. By considering a Gedanken experiment 
in which some hot gas is thrown into a black hole, J.~Bekenstein conjectured 
\cite{Bekenstein:1973ur} that a black hole should itself  have an entropy proportional 
to its horizon area $S \propto A$ in order that the second law of thermodynamics 
is not violated. Shortly after this, S.~Hawking \cite{Hawking:1974sw} showed, by 
studying a quantum field theory on a classical black hole space-time, that black 
holes will actually emit  thermal radiation with a temperature 
$T= \hbar \frac{\kappa}{2 \pi}$. Thus, identifying the heat flow of some microscopic 
degrees of freedom at a temperature $\hbar \frac{\kappa}{2 \pi}$ crossing the horizon to be 
\beq \label{clheat}
\delta Q = \delta M - \Omega\, \delta J -  \Phi\, e\, \delta q
\eeq
the first law of black hole thermodynamics implies that the entropy of the black hole is given by \eq{entropy}.
From now on we will use units $\hbar =1$. S.~Hawking's original derivation 
of the black hole entropy \eq{entropy} centrally relied on a thermodynamical reasoning 
and a semi-classical approximation for quantum gravity. Subsequently it was 
shown by G.~Gibbons and S.~Hawking \cite{Gibbons:1976ue} that it is also possible 
to obtain these results directly from the Euclidean path integral for quantum gravity by
taking the 
Einstein-Hilbert action (with vanishing cosmological constant and 
appropriate boundary terms) as the saddle point
approximation. Thus, the entropy \eq{entropy} also corresponds 
to the correct statistical entropy within this approximation to the full path integral.

\subsection{Coarse-graining and the renormalisation group}
The renormalisation group  is a powerful tool to study  the scale- or 
energy-dependence of quantum field theories and statistical systems.
The essential idea in the construction of non-perturbative renormalization 
group equations, as put forward by L.~Kadanoff and K.~Wilson \cite{Wilson:1973jj, Wilson:1975ly}, 
is  to integrate-out the  short-distance fluctuations, gradually, ordered according 
to their characteristic energy by means of a momentum cut-off $k$.  As such, 
Wilson's approach leads to a coarse-grained version of quantum field theory 
which continuously interpolates between the micro- and macrophysics \cite{Berges:2000ew}. 
In modern formulations, the coarse-graining  is 
achieved by adding an infrared cutoff $R_k$ at momentum scale $k$ to the propagators which, 
 within a few constraints, can be chosen freely \cite{Litim:2000ci,Litim:2001up,Litim:2002qn}.
The effect of this procedure is to generate a scale-dependent or 
``flowing" effective action $\Gamma_k$  \cite{Wetterich:1989xg}
which interpolates between a microscopic action at large RG scale $(k\to\infty)$, 
and the full quantum effective action in the long-distance limit $(k\to 0)$ \cite{Berges:2000ew}.  
Most importantly, the effective action obeys an exact functional identity \cite{Wetterich:1992yh}
\begin{equation}
\label{FRG}
\partial_t\Gamma_k=\frac12\Tr\left(\Gamma^{(2)}_k+R_k\right)^{-1}\partial_t\,R_k
\end{equation}
which relates an infinitesimal change of $\Gamma_k$ at RG scale $t=\ln k$  to a momentum 
trace over the propagator.  For  a simple mass term  $R_k\propto k^2$ the RG flow \eq{FRG} 
reduces to the well-known Callan-Symanzik equation.  By construction, the inverse of the RG 
momentum scale is interpreted as the linear ``resolution"
\begin{equation}\label{resolution}
\ell\approx 1/k_{\rm phys}
\end{equation}
of the RG ``microscope" at which the physics is observed \cite{Wetterich:1989xg,Berges:2000ew}. 
Ultimately this is a consequence of the RG flow \eq{FRG} being  local in field- and 
momentum-space, meaning that the change of the effective action at energy scale $k$ 
is induced by the fluctuations of the quantum fields at about that energy scale.

For gravity, these ideas have been put forward in
\cite{Floreanini:1993na,Reuter:1996cp,Litim:2003vp} (see \cite{Litim:2006dx,Niedermaier:2006ns,Niedermaier:2006wt,Percacci:2007sz,Litim:2008tt,Litim:2011cp,Reuter:2012id} for reviews), primarily to provide a computational framework within which S.~Weinberg's asymptotic safety conjecture for gravity can be addressed \cite{Weinberg:1980gg}. The set of scale-dependent 
gravitational actions $\Gamma_k[g_{\mu\nu}]$ then describes the ``quantum spacetime" 
obtained from integrating-out gravitational fluctuations down to the energy scale $k$. 
The coarse-grained metric field $\langle g_{\mu\nu}\rangle_k$, which solves the 
effective equations of motion 
\begin{equation}
\frac{\delta\Gamma_k}{\delta g_{\mu\nu}}=0\,,
\end{equation}
has the interpretation of a Riemannian manifold averaged over the length scale 
\eq{resolution} \cite{Lauscher:2005qz,Reuter:2007fk,Reuter:2011ah}. The RG trajectory $k\to \Gamma_k$ 
encodes how the appearance of the physical system, characterised by  the effective metric field 
$\langle g_{\mu\nu}\rangle_k$, changes with the resolution of the RG ``microscope" 
\cite{Lauscher:2005qz}.

Many applications of the gravitational RG flow have  dealt with the search for 
ultra-violet (UV) fixed points, a prerequisite for metric gravity to become a well-defined 
local quantum field theory at high energies following the asymptotic safety 
conjecture \cite{Weinberg:1980gg,Litim:2006dx,Niedermaier:2006ns,Niedermaier:2006wt}. 
By now, evidences for UV fixed points in gravity 
have been found in four-dimensional Einstein-Hilbert gravity 
\cite{Reuter:1996cp, Souma:1999at, Lauscher:2002vn, Litim:2003vp,Niedermaier:2009zz,Donkin:2012ud,Christiansen:2012rx},  
higher-dimensional gravity \cite{Litim:2003vp,Fischer:2006fz,Fischer:2006at}, 
higher-derivative gravity \cite{Codello:2008vh, Benedetti:2009gn,Falls:2013bv}, and coupling to 
matter fields \cite{Percacci:2002ie,Percacci:2003jz,Daum:2009dn,Folkerts:2011jz}. A gravitational 
fixed point is also consistent with results from holography \cite{Litim:2011qf,Litim:2012zz}, conformal reductions \cite{Reuter:2008wj}, Lorentzian signature \cite{Manrique:2011jc}, and minisuperspace approximations \cite{Litim:2012vz}.
Phenomenological implications of a gravitational fixed point have been explored 
for black holes 
\cite{Bonanno:2000ep, Bonanno:2006eu, Falls:2010he, Cai:2010zh, Burschil:2010ys,Reuter:2011uq,Becker:2012js},  
cosmology 
\cite{Bonanno:2002fk, Bonanno:2002uq, Weinberg:2009wa, Bonanno:2010bt, Koch:2010nn,Bonanno:2010mk, Contillo:2010ju,  Tye:2010an, Hindmarsh:2011hx, Contillo:2011fn,Hong:2011ws, Bonanno:2011kx,Hindmarsh:2012rc}, and particle physics 
\cite{Litim:2007iu,Hewett:2007st,Gerwick:2011jw,Gerwick:2010kq,Dobrich:2012nv}. 
Non-local low-energy corrections to the gravitational effective action have equally been 
addressed \cite{Machado:2007ea,Satz:2010uu}. 

In the remaining part of this paper, we adopt the RG and the 
scale-dependent gravitational action $\Gamma_k$ to develop an RG improved version to the 
laws of black hole thermodynamics.

\section{Black holes under the microscope} \label{secII}

In this section we introduce our set-up to implement quantum corrections to the 
thermodynamics of black holes using a continuous Wilsonian renormalisation group.

\subsection{Action}
We are interested in a four-dimensional theory involving gravity, 
$U(1)$ gauge fields, and possibly matter fields. In the spirit of a scale-dependent 
effective action we describe their dynamics in terms of the ``flowing" Einstein-Hilbert 
action coupled to photons and matter, approximated by
\beq \label{action}
\Gamma_k[g_{\mu \nu},A_\mu] =  
\int d^4x \sqrt{-\det g_{\rm \mu\nu}} \left[ \frac{1}{8 \pi G_k} {\cal R} 
+ \frac{1}{4 \alpha_k} F^{\mu \nu} F_{\mu \nu} \right]+S_{m}\,.
\eeq
Here,  ${\cal R}$ denotes the Ricci scalar and $F$ the field strength of the 
photon, and $S_{m}$ stands for a possibly scale-dependent matter action.
The effective action differs from the classical Einstein-Hilbert action coupled to 
matter in that all couplings are considered as running couplings whose tree level 
approximation describes the quantum effects of modes down to the energy scale $k$.
It  is understood as a  solution to the RG flow for the Einstein-Maxwell theory in its 
domain of validity.  In the deep infrared limit where the RG scale is removed $(k\to 0)$ 
both the running Newton coupling $G_k$ and the running fine structure constant 
$\alpha_k\equiv {e_k^2}/{(4 \pi)}$ will approach their low-energy values 
$G\approx 6.674 \times 10^{-11} {\rm \ N}\, {\rm (m/kg)^2}$ and $\alpha\approx \frac{1}{137}$. 
We assume that the scale-dependence of Newton's coupling $G_k$ and of the fine structure constant 
are known, at least approximatively, though the actual form of these functions is not important for our line of reasoning. 

For large $k$, we will approach a fine grained action for high momentum modes. 
In perturbative quantum gravity, the action \eq{action} would then cease to be a good 
approximation due to the non-renormalisability of gravity.
On the other hand, if metric quantum gravity becomes asymptotically safe, the action 
persist towards higher energies \cite{Reuter:1996cp, Souma:1999at, Lauscher:2002vn, Litim:2003vp,Niedermaier:2009zz}. The RG flow of euclidean Einstein-Hilbert gravity coupled to a $U(1)$ 
gauge field has recently been considered in \cite{Harst:2011zx}.

\subsection{Black holes and entropy}
At fixed $k$,  and by varying $\Gamma_k$ with respect to the metric and the gauge 
fields we recover the Einstein-Maxwell theory coupled to an energy momentum tensor 
$T_m^{\mu \nu}$ and a current $J^{\mu}$ obtained from the matter action $S_{\rm m}$. 
Setting $J^{\mu} =0$ and $T_m^{\mu \nu}=0$ Kerr--Newman-type black holes are the 
unique stationary black hole solutions. The sole difference with the standard solutions 
is that the couplings $G_k$ and $\alpha_k$ explicitly take $k$-dependent values. 
As such we have a family of  Kerr-Newman black hole solutions characterised by a  
fundamental relation between its mass $M$, the horizon area $A$, charge $q$, and 
angular momentum $J$, and the RG scale $k$. This relation has the form 
\beq \label{areaonshell}
A= A(M,J,q;k)
\eeq
where the scale-dependence enters the equation only implicitly via the couplings 
$G_k$ and $e^2_k$. The equation \eq{areaonshell} expresses an on-shell relation 
with respect to the underlying action $\Gamma_k$.  The scale $k$ indicates that 
degrees of freedom with momenta above $k$ have been integrated out to give 
rise to a semi-classical space-time geometry. It is our assumption that these 
microscopic degrees of freedom also give rise to the thermodynamical properties 
of space-time. Under this assumption we think of their black hole entropy
\beq \label{Sk}
S_k= \frac{A}{4 G_k}
\eeq
as accounting for those degrees of freedom which have already been integrated out 
from the path integral. It is worth noting the parametric dependence of \eq{Sk} on $G_k$, which states that the entropy per area increases with decreasing gravitational coupling $G_k\to 0$, and vice-versa.
With the area $A$ given by \eq{areaonshell} the relation \eq{Sk} 
will give an on-shell expression for the entropy $S_k= S_k(M,J,q;k)$. We could also 
consider an off-shell definition for the  entropy where it is  not assumed that the area 
is given by  \eq{areaonshell}, but instead take \eq{Sk} as the Wald entropy  
\cite{Wald:1993zr} obtained from the underlying action  \eq{action}. Consequently 
the entropy would depend  on the metric, via $A$ and, independently, on the scale $k$.
The RG flow for the off-shell entropy \eq{Sk}  taken at constant area is then given by
\beq \label{Sdk}
\frac{\partial }{\partial \ln k} S_k = -S_k \frac{\partial \ln G_k}{\partial \ln k} \, 
\eeq
and  only depends on the RG flow of $G_k$, and not on the on-shell relation \eq{areaonshell}. 
We can think of this flow for the entropy as the ``focusing of the microscope'' through which 
the physics is viewed, in contrast  to a change of the underlying state of the system which
would additionally lead to a variation of the area $\delta A$.
The family of Kerr-Newman black holes with \eq{areaonshell} obeys the standard laws 
of black hole thermodynamics for all $k$. 
This is so because the thermodynamical nature 
of  black hole solutions to \eq{action} is independent of the actual numerical values of the couplings. 
These relations are modified as soon as the RG scale $k$ is linked to the physical parameters 
of the black hole solution, to which we turn next.

\subsection{Scale identification}

In order to develop a renormalisation group improved version of black hole thermodynamics, 
we identify the degrees of freedom responsible for the thermodynamical properties of the black 
hole with those that have been integrated out in the underlying path integral. To this end we 
adjust the RG scale to the physical parameters of the black hole. For asymptotically large black holes the effective action approaches the classical infrared limit, and the relevant RG scale becomes very small $k\to 0$. For finite-size black holes, lesser modes are required in the underlying path integral to constitute the background geometry. The relevant RG scale at which to evaluate the effective action \eq{action} should then be finite $k>0$, and the value of the running couplings may be different from their infrared values.  
We will thus assume that there exists an ``optimal" RG scale $k=k_{\rm opt}$ at which to evaluate the couplings, set by  the macroscopic spacetime geometry with black hole parameters $M$, $J$, and $q$, 
\beq\label{kopt}
k=k_{\rm opt}(M,J,q) \, .
\eeq
Heuristically, if the RG scale is taken much larger than $k_{\rm opt}$, the effective action $\Gamma_k$ is not yet a good tree level approximation for a black hole solution 
with physical parameters $M$, $J$ and $q$, 
and additional quantum (loop) corrections will have to be taken into account. 
On the other hand, for $k$ much smaller than $k_{\rm opt}$ the effective action and its saddle point solution may become 
too coarse-grained.\footnote{
This line of reasoning is similar to an optimized scale identification used in the 
context of inflation \cite{Weinberg:2009wa}.}  
Under this assumption we will again have a set of Kerr-Newman-type black holes parameterised 
by $M$, $J$ and $q$, except that now the space of black hole solutions
is deformed by the underlying RG trajectory through the link \eq{kopt}. 
As a result, a new state function 
\begin{equation}\label{statefunction}
A=A(M,J,q)
\end{equation} 
is obtained by inserting $k=k_{\rm opt}(M,J,q)$  into \eq{areaonshell} which, in general,  
may be different from the classical state function. Below we show that the scale $k_{\rm opt}(M,J,q)$ 
is  fixed up to an overall normalisation, provided that the black holes obey a scale-dependent version 
of black hole thermodynamics. In order to achieve this goal we must decide on the appropriate 
generalisation for the variation of the entropy $\delta S$. We will take this variation as 
\beq \label{deltaS}
\delta S_{k_{\rm opt}} = \frac{\delta A}{4 G_{k_{\rm opt}}}\,.
\eeq
Note that this corresponds to taking the partial derivative with respect to $A$. This choice amounts to a variation of the off-shell entropy with respect to the metric field at 
fixed RG scale $k$. This is similar to how the equations of motion are obtained from $\Gamma_k$, 
and ensures that we compare entropies which are defined with respect to the same coarse-graining 
scale. However it is important to note that the form of the second law will be changed due to the $k$-dependence of the couplings appearing in $A(M,J,q;k)$. Below we will explicitly show that this leads to additional RG-induced terms in $\delta A$ which provide corrections to the semi-classical Hawking temperature, see \eq{implicit}.  If, on the other hand, we are taking the full exterior derivative of $S_{k_{\rm opt}}$ we would 
instead gain an extra term originating from the flow \eq{Sdk}, giving
\beq \label{deltaSalt}
\delta S_{k_{\rm opt}}  =  \frac{\delta A}{4G_{k_{\rm opt}} }- S_{k_{\rm opt}}  \delta \ln G_{k_{\rm opt}}  \,.
\eeq
In addition to the corrections arising implicitly through $\delta A$, this form of the entropy also contains corrections explicitly proportional to the variation of $G_k$.  The interpretation of \eq{deltaSalt} is that it compares two different entropies defined relative to two 
distinct coarse-graining scales. In the spirit of our construction, we will therefore take \eq{deltaS} in favour of 
\eq{deltaSalt}. 
Below, we will provide additional arguments related to conformal scaling in the UV limit  (Sec.~\ref{Temperature}), and the equivalence with findings from  RG-improved metrics (Sec.~\ref{thermo})  to further strengthen the choice made here.

\subsection{ Thermal equilibrium}
Next we determine the scale \eq{kopt} entering the relation \eq{areaonshell} using a 
thermodynamical bootstrap. Assuming that \eq{kopt} is given as a function of $M$, $J$ 
and $q$ we  perform a Gedanken experiment and allow a small amount of matter to fall into 
the black hole which thereby will change in mass, charge, and angular momentum to settle down 
into a new state corresponding to the mass $M+ \delta M$, angular momentum $J+ \delta J$ 
and charge $q+ \delta q$. This process 
induces a change in the scale \eq{kopt}  into  $k_{\rm opt}+\delta k_{\rm opt}$. 
In order to describe this process thermodynamically we have to relate the change in heat with the 
change in entropy. We will assume that the relation
\beq  \label{state}
\frac{\delta Q}{T} = \delta S_{k_{\rm opt}} 
\eeq
holds true, with the variation in entropy taken as  \eq{deltaS}. In the light of the results by 
T.~Jacobson  \cite{Jacobson:1995ab}, the equation \eq{state} has a natural interpretation as a 
RG improved form of Einstein's equations on  the black hole horizon.  In addition, and on general thermodynamical 
grounds we expect that a thermal description of the black hole embodied by the relation \eq{state}  
should be valid provided the entropy and the specific heat are large  \cite{Preskill:1991tb},
\beq\label{validity}
\begin{array}{ccl}\displaystyle
{1}/{S}&\ll& 1 \\[2ex]
\displaystyle
\left| 
\frac
{\partial T}{\partial M} \right|_{J,q} &\ll&1 \,.
\end{array}
\eeq
 We now turn to the heat $\delta Q$  crossing the horizon which is given by
\beq \label{heat}
\delta Q =  \delta M - \Omega\, \delta J - \Phi\, e_{k_{\rm opt}} \, \delta q \,.
\eeq
The heat is understood as the energy carried by the coarse-grained degrees of freedom 
with energy larger than \eq{kopt}.  These are the degrees of freedom that have been 
integrated out in the path integral to obtain the effective equations of motion, in analogy to 
the ``integrating-out" of individual atoms or molecules which carry heat in a standard thermodynamical 
description of a gas. To continue, we note that the total change in the area  of the black hole is given by
\beq \label{dA}
\delta A = 4G_{k_{\rm{opt}}} \frac{2 \pi}{\kappa} \delta Q 
+\left. \frac{\partial A(M,J,q;k)}{\partial \ln k}\right|_{k=k_{\rm{opt}}} 
 \frac{\delta k_{\rm{opt}}}{k_{\rm{opt}}}\, .
\eeq
The first  term follows from \eq{firstlaw} since at constant $k$ we  obtain the classical 
variation of the area. The second term takes the implicit scale-dependence of $A$ into 
account. It is proportional to the RG $\beta$-functions of the couplings
 and therefore accounts for the quantum corrections. In the approximation  considered here \eq{action}, the scale-variation of the area in \eq{dA} contains two terms originating from the scale-dependence of the fine structure and Newton constants. Recalling \eq{areaonshell}, which states that the $k$-dependence of the area arises via the $k$-dependence of couplings $A(M,J,q;k)=A(M,J,q; G_k ,\alpha_k)$, the additional terms read explicitly
 \beq\label{implicit}
 \left. \frac{\partial A(M,J,q;k)}{\partial \ln k}\right|_{k=k_{\rm{opt}}} 
 \frac{\delta k_{\rm{opt}}}{k_{\rm{opt}}} =  \left.\left( \frac{\partial A}{\partial G_k} \frac{\partial G_k}{\partial \ln k} + \frac{\partial A}{\partial \alpha_k} \frac{\partial \alpha_k}{\partial \ln k} \right)\right|_{k=k_{\rm{opt}}} 
 \frac{\delta k_{\rm{opt}}}{k_{\rm{opt}}} \,.
 \eeq
The appearance  of the new terms \eq{implicit}
 imply that we go off-shell with respect to the equations of motion at scale $k_{\rm opt}$ to 
 obtain a solution to the equations at a scale $k_{\rm opt} + \delta k_{\rm opt} $.
In order to identify the scale $k_{\rm opt}$ which appears in \eq{dA} we rearrange this 
expression for $\delta Q$ and insert it into the RHS of \eq{state}. With the LHS given 
by \eq{deltaS} we obtain the relation
\beq \label{deltaA}
\left(1 - \frac{2 \pi}{\kappa} T \right) \delta A = 
\left. \frac{\partial A(M,J,q;k)}{\partial \ln k}\right|_{k=k_{\rm{opt}}} 
\frac{ \delta k_{\rm{opt}}}{k_{\rm{opt}}}\,.
\eeq
The significance of \eq{deltaA} is as follows. The classical relation between temperature 
and surface gravity $T =\frac{\kappa}{2 \pi}$ holds true provided the RHS vanishes.
In the presence of RG corrections, the RHS 
describes  corrections to the temperature of the black hole. Most importantly, 
we note that $\delta k_{\rm{opt}}$  
must be proportional to  $\delta A$ independently of the specific form for the heat $\delta Q$. 
This implies that the scale $k_{\rm{opt}}$ depends on $M$, $J$ and $q$ only through the combination 
\beq\label{kA}
k_{\rm{opt}}(M,J,q)\equiv k_{\rm{opt}}(A(M,J,q))\,.
\eeq
Thus we are 
lead to the conclusion, via a thermodynamical argument, that the black hole area $A$ is the 
unique scale associated to the black hole geometry which determines the renormalisation 
group scale $k_{\rm{opt}}(A)$. Dimensional analysis
then dictates that this 
relation reads
\beq \label{k}
k_{\rm{opt}}^2 = \frac{4 \pi}{A}\,\xi^2
\eeq
where the factor $4\pi$, the surface of the unit $2$-sphere, is conventional and $\xi$ is an 
undetermined dimensionless constant. The scale identification \eq{k} has a straightforward 
generalization to dimensions different from four.  

The thermodynamical bootstrap fixes the 
relation \eq{k} only up to an unknown proportionality factor. This is reminiscent of the standard 
laws of black hole thermodynamics being independent of the numerical values of couplings.  
The occurrence of the factor $\xi$ is understood from the RG point of view as the freedom of 
choosing the normalisation for $k$, which comes about via the Wilsonian momentum cutoff 
$R_k$. Hence the coefficient $\xi=\xi(R_k)$ depends on the RG scheme inasmuch as the 
value of $k_{\rm opt}=k_{\rm opt}(R_k)$ depends on it,  to ensure that the effective physical 
cutoff scale  $k_{\rm phys}\approx k_{\rm opt}/\xi$ is scheme-independent.  For physical choices of 
the RG scheme we expect $\xi$ to be of order unity, and assuming that this has been done 
we will set $\xi=1$  for the remaining part of the paper.

In this light, the result \eq{k}  states that the underlying effective action $\Gamma_k$, 
\eq{action}, should be evaluated at the RG scale $k_{\rm opt}$  set by the horizon area of 
its black hole solution. In particular, since quantum fluctuations of momentum modes larger 
than $k_{\rm opt}$ have indeed been integrated out, the black hole area acts as a 
diffeomorphism invariant infrared cutoff for its effective action. This result is  consistent with 
the view that  thermodynamic properties  originate from those degrees of freedom which 
constitute the black hole.  

\subsection{RG thermodynamics}
We are now in a position to define the renormalisation group improved relation \eq{statefunction} between the area $A$
and the physical parameters $M$, $J$ and $q$ by replacing the classical couplings 
by running couplings evaluated at the scale \eq{k}. This is most neatly expressed in terms of a mass function $M=M(A,J,q)$, with
\bea \label{Mimp}
M^2 & \equiv & \frac{4 \pi}{A} 
\left[\left(  \frac{A+ 4 \pi G_{\rm opt}(A) e_{\rm opt}^2(A) q^2}{8 \pi G_{\rm opt}(A) }\right)^2 + J^2 \right]\,.
\eea
It defines initial and final states of a thermodynamical process, in conjunction 
 with a small RG transformation. The mass function is obtained from the standard relation 
 for the Kerr-Newman black hole by replacing the classical coupings with 
 $G_N\to G_{\rm opt}(A) \equiv G_{k_{\rm opt}(A)}$ and $e^2\to e_{\rm opt}^2(A)\equiv e^2_{k_{\rm opt}(A)}$ 
 under the identification \eq{k}. The relation \eq{Mimp} then allows us to parameterise 
 these states simply by the mass $M$, charge $q$, and angular momentum $J$ thus
recovering a RG improved version of black hole uniqueness. Solving for $A$ we find  
RG improved state functions  $A(M,J,q)$. If there are several  roots $A_i$ for the same 
values of $M$, $J$ and $q$ these have the natural interpretation as multiple horizons 
for the same black hole e.g.~inner and outer horizons of a Kerr black hole as in the 
classical theory. Note that since these horizons generically have different entropies and 
temperatures, being in thermal equilibrium with either of them  corresponds to a different 
thermodynamical state. Their  entropy is given by
\beq \label{Simp}
S_{k_{\rm opt}}  = \frac{A}{4G_{\rm opt}(A)}
\eeq
with its thermodynamical variation given by \eq{deltaS}. At this point it is useful to remember 
that the scale $k$ tells us which degrees of freedom have been integrated out in the path 
integral and that the relation \eq{k} is obtained by requiring that $k$ is optimised according 
to the background geometry. So the entropy \eq{Simp} counts the number of degrees of freedom 
that have been integrated out in this optimal coarse graining.

The temperature $T$, angular velocity $\Omega$ and electric potential $\Phi$ appear in an 
improved first law of black hole thermodynamics obtained by putting the variation of the 
entropy \eq{deltaS} on the RHS of \eq{state} and \eq{heat} on the LHS, leading to
\beq \label{improvedfirstlaw}
T \,  \frac{\delta A}{4G_{\rm opt}(A)} =   \delta M - \Omega \delta J - \Phi e_{\rm opt}(A)  \delta q \, .
\eeq 
This differs from the standard first law by the presence of the area-dependent couplings. 
Also,  the relation between temperature and the classical expression for the surface gravity 
of the black hole receives RG corrections. (We will see in Sec.~\ref{secIV} that there exist 
explicit RG-improved metrics for which \eq{improvedfirstlaw} holds true with the temperature 
identified with the surface gravity felt by a test particle on these black hole metrics.)
The intrinsic quantities  $T$, $\Omega$ and $\Phi$ are obtained by taking derivatives of 
$M$ (or  $A$) in line with \eq{improvedfirstlaw}. The RG improved black hole temperature is 
obtained as
\beq \label{T}
\frac{1}{T} =\frac{1}{4G_{\rm opt}(A)} \frac{\partial A}{\partial M}  
\eeq
which receives corrections containing derivatives of the couplings and their RG 
$\beta$-functions. On the other hand both $\Omega$ and  $\Phi$ can be simply obtained 
from their classical expressions by replacing the classical couplings by the functions 
$e_{\rm opt}(A)$ and $G_{\rm opt}(A)$. This `factorization' holds true since derivatives of 
\eq{Mimp} with respect to $J$ and $q$, by the virtue of \eq{k}, cannot touch the running 
couplings as they only depend on the area $A$.

Provided we use \eq{deltaSalt} on the RHS of \eq{state}, rather than \eq{deltaS} as we did, the 
derivation of \eq{kA} and \eq{k} remains unaffected.
However, we would obtain a different expression for the temperature, replacing \eq{T} by 
\beq\label{T2}
\frac{1}{T} = \frac{1}{4G_{\rm opt}(A)} \frac{\partial A}{\partial M} 
 \left(1- \frac{\partial \ln G_{\rm opt}(A)}{\partial \ln A} \right) \,.
\eeq
 In particular  this would imply that if $G(A) \propto A$, the temperature would diverge 
 due to the vanishing of the bracket on the RHS. In this paper we always take \eq{deltaS} to 
 define the variation of the entropy  leading  to \eq{T}, where no such divergence appears.

 At a practical level the formalism  presented here allows us to obtain models 
 of quantum black hole thermodynamics given an RG trajectory for $G_k$ and 
 $e_k$. This provides  a controlled way to include quantum corrections without 
 moving too far from the semi-classical thermodynamics of black holes. Ultimately 
 such a (thermal) description may break down at high energies where we expect 
 that the action \eq{action} should include higher order terms and where the thermodynamical 
 approximation based on Kerr-Newman black holes will no longer be a good one.

\subsection{Semi-classical limit}
 For low energies we must recover classical general relativity such that astrophysical 
 black holes are described by the Einstein-Maxwell equations. This is achieved provided 
 we have an RG trajectory with the limits
 \begin{equation}
\begin{array}{ccl} 
G_k \to G_N&\quad {\rm for}\quad & k \ll M_{P}  \\[1ex]
 e^2_k \to e^2& {\rm for} & k \ll m_e
\end{array}
\end{equation}
where $M_{P}= 1/\sqrt{G_N}$ is the Planck mass and $m_e$ is the electron mass.
For Newton's coupling, the limit of classical general relativity is achieved as a 
consequence of IR attractive fixed points. The scale identification \eq{k} implies that these 
limits are achieved for a black hole with a sufficiently large area $A$ as this entails that the 
underlying effective action is integrated down to $k\to 0$. Consequently, astrophysical 
black holes will then be described accurately by classical general relativity. Our model 
of RG improved black hole thermodynamics then passes the first mandatory test of 
recovering the right semi-classical limit in the infra-red.
We note that since $M_{P} \gg m_e$ there exists a large range of scales for which 
gravity remains semi-classical, but where the running of $e_k$ will induce quantum 
corrections to tiny charged black hole as soon as the radius of the black hole 
approaches the Compton wavelength of an electron.  

 \section{Thermodynamics and asymptotic safety}\label{secIII}

Our reasoning in the previous sections was  independent of the actual form of the 
running couplings $G_k$ and $\alpha_k$ and, therefore, of the UV completion of gravity. 
In this section we consider an explicit example where gravity becomes anti-screening at short 
distances as predicted by the  asymptotic safety conjecture for gravity  \cite{Weinberg:1980gg}.

 \subsection{Fixed point and characteristic energy}\label{FP}
 Asymptotic safety is a non-perturbative generalisation of asymptotic freedom for which the 
 relevant couplings of a theory reach a non-Gaussian fixed point at high energies. If realised 
 in Nature, the asymptotic safety conjecture implies that the short-distance  fluctuations of 
 gravity shield the theory from the divergences of standard perturbation theory   \cite{Weinberg:1980gg}. 
  Furthermore, to ensure predictivity, the number of relevant directions flowing away from the 
  fixed point must be finite. There has recently been much evidence that this is the case of gravity 
  \cite{Reuter:1996cp, Souma:1999at, Lauscher:2002vn, Litim:2003vp, Codello:2008vh, Benedetti:2009gn}. 
For Newton's constant this implies that the dimensionless coupling $g_k= k^2 G_k$ reaches a fixed point
 $g_k \to g_* \neq 0$ in the UV limit, implying near-conformal behaviour with a characteristic weakening $G_k \to g_*/k^2$ at short distances. The 
 presence of such an UV fixed point also implies that gravity may exist as a QFT to arbitrarily 
 large energy scales and is therefore non-perturbatively renormalisable.

To explore the implications of  the asymptotic safety
conjecture for the physics of black holes we 
allow for a non-trivial scale-dependence of Newton's constant. 
In terms of the graviton anomalous dimension $\eta=\frac{d\ln G_k}{d\ln k}$, the 
RG flow reads \cite{Litim:2006dx,Niedermaier:2006ns,Niedermaier:2006wt}
\beq\label{dG}
\frac{d\,G_k}{d\ln k}=\eta\,G_k\,.
\eeq
In general, the anomalous dimension is a function of all couplings 
of the theory. In perturbation theory, one finds
$\eta=-2\,\omega\,k^2 G_k+{\cal O}(G_k^2)$
where the sign of the one-loop coefficient $\omega$ depends on the field 
content of the theory. Gravity is perturbatively anti-screening if $\omega$ is positive. 
In the IR limit, the anomalous dimension and \eq{dG} are arbitrarily small 
meaning that $G_k\approx G_N$. At a non-trivial fixed point the anomalous 
dimension becomes large, $\eta = -2$, to ensure that the dimensionless 
gravitational coupling $G_k\,k^2$ approaches a non-trivial UV fixed point 
of gravity  $g_*$. Analytical RG flows which interpolate between these 
limit have been given in \cite{Litim:2003vp}. 
For our purposes, a good approximation for the integrated RG flow is given by
\beq \label{Gk}
\frac1{G_k} = \frac1{G_N} + \frac{k^2}{g_*}\,.
\eeq
In the infrared limit the running coupling reduces to its classical value. In the UV limit 
the second term takes over leading to the asymptotic weakening of gravity $G_k \to 0$. 
Note that $g_*$ plays a double r{$\hat {\rm o}$le in the RG flow \eq{Gk}. In the IR limit, $1/g_*$ represents 
the perturbative one-loop coefficient $\omega$. In the UV limit, $g_*$ stands for the non-perturbative 
fixed point. In the full theory, these numbers can be different from each other.
Typically one finds values $g_*$ of order unity. For the purpose of this study, we shall keep $g_*$ 
as a free parameter. Also, the flow \eq{Gk}  is non-perturbatively anti-screening 
as long as $g_*$ is positive, in agreement with RG results for purely gravitational flows. 
In set-ups where the RG running of Newton's coupling is dominated externally, eg.~by 
matter fields, the one-loop coefficient may turn negative. Returning to \eq{Gk}, we note that the 
quantum corrections are responsible for the appearance of a characteristic energy scale 
\begin{equation}\label{kc}
E^2_c={g_*}\,{M^2_P}
\end{equation}
where we have introduced the Planck mass $M_P$, with $M_P^2 \equiv 1/{G_N}$. At the 
energy scale $k=E_c$ we have that the tree level term equals the quantum corrections in 
magnitude, and hence the scale $E_c$ sets the boundary between IR and UV scaling. We 
also note that the quantum corrections are suppressed in the limit where $1/g_*\to 0$. The 
meaning of this limit is that the theory still owns an UV fixed point except that it is infinitely 
far away and cannot be approached within finite RG `time' $t=\ln k$. This is equivalent to a semi-classical 
approximation with no RG running at all, corresponding to the limit $\hbar\to 0$.

\subsection{Critical mass and area}
We now show that an asymptotically safe RG running such as \eq{Gk} with the cross-over 
scale \eq{kc},  in conjunction with the reasoning of the previous section, lead to the 
appearance of a new mass scale
\beq\label{Mc}
M_c^2 = \frac{1}{g_*} M_P^2 \,.
\eeq
The scale $M_c$ owes its existence to the presence of  the fixed point $g_*$ and is hence 
absent in the  classical theory. It is qualitatively different from the classical Planck scale $M_P$ which is an 
infrared parameter of  Einstein gravity. Note also that $M_c$ is dual to the 
energy scale $E_c$,
\begin{equation}
\label{dual}
M_c\,E_c=M^2_P
\end{equation}
irrespective of the value for $g_*$. Extended fixed point searches in pure four-dimensional quantum gravity indicate that $g_*$ is of the order one, with $M_c\approx E_c$ and $M_c\approx M_P$. The classical limit is 
recovered by taking  $1/g_* \to 0$ where the mass  scale $M_c \to 0$ disappears. 

The significance of the mass scale \eq{Mc} can be understood from the following observations. (For simplicity 
we restrict the discussion to the case where $q=0$.)
 We insert the running coupling \eq{Gk} into \eq{Mimp} to find
\beq \label{MAS}
M^2 = \frac{4 \pi}{A} \left( \frac{(A + 4 \pi G_N^2 M_c^2)^2}{64 \pi^2 G_N^2} + J^2 \right)\,.
\eeq
This function encodes all the relevant information needed to obtain properties 
of the RG improved black hole via the first law \eq{improvedfirstlaw}. Note that 
it takes a  form similar to the classical Kerr-Newman black hole (i.e.~\eq{Mimp} 
with constant $G$ and $e$) with $M_c^2/M_P^2$ playing the role of the classical 
charge $(e\,q)^2$. Taking the limit $M_c \to 0$ we obtain the classical Kerr black 
hole relation between the mass, area and angular momentum.
Leaving $M_c$ non-zero we can solve \eq{MAS} to find the quantum-corrected area 
$A_\pm(M,J)$ of the outer and inner horizons of the RG improved black hole, 
\begin{eqnarray} \label{ASarea}
 A_\pm = 4 \pi G_N \left( 2G_N M^2  -G_N M_c^2   
 \pm 2 \sqrt{G_N^2 M^4 - J^2  - G_N^2 M_c^2 M^2 }  \right) \,.
\end{eqnarray}
Taking a derivative of this expression with respect to the mass $M$ one can find 
the temperature of the black hole $T$ from the first law \eq{improvedfirstlaw}. 
Similarly one may find the angular momentum by taking a derivative with respect 
to $J$. When the expression inside the square root of \eq{ASarea} vanishes we 
have degeneracy between inner and outer horizons $A_+ = A_-$ and the temperature 
of the black hole falls to zero. This corresponds to an extremal black holes with mass
 \beq\label{Mex}
 M_{\rm ex}(J)^2 = \frac{1}{2}  \left(M_c^2 
 + \sqrt{4 J^2 + \left(\frac{M_c}{M_P}\right)^4}M_P^2\right) \,.
\eeq
In the classical limit the extremal black hole mass \eq{Mex} reduces to the extremal Kerr mass 
\beq\label{MKerr}
M_{\rm Kerr}^2(J) = 
|J|\ M_P^2\,.
\eeq  
The physical meaning of the  mass scale \eq{Mc}  
then follows from \eq{Mex} in that it characterizes the  mass of the smallest achievable 
black hole $M_c=M_{\rm ex}(J=0)$ with a causal horizon. Here, the existence of a lightest  
black hole is a direct 
consequence of the RG equations for $G_k$. As we probe gravity at smaller distances 
the anti-screening effects weaken the gravitational interactions such that a black hole 
horizon can no longer form, and the notion of a semi-classical black hole space-time 
ceases to exist. The horizon area of the smallest black hole is given by 
$A_c = 4\pi (G_N\,M_c)^2$, which can be written as
   \beq\label{Ac}
 A_c = \frac{4 \pi}{g_* M_P^2}\,.
 \eeq
Using \eq{Ac} together with \eq{k} identifies the  RG scale corresponding to the 
smallest black hole as the cross-over scale \eq{kc}, $k_{\rm opt}= E_c$.
We also note that for masses $M>M_c$ and vanishing angular momentum $J=0$ an 
inner horizon of area $A_-$ will always be present. This holds true independently of the 
detailed form of the RG equation \eq{Gk}, showing that the degeneracy of the
Schwarzschild black hole, which classically does not display a Cauchy horizon, 
is lifted by asymptotically safe quantum gravity fluctuations.  
   \begin{figure*}[t]
\begin{center}
\unitlength0.001\hsize
\begin{picture}(800,550)
\put(0,0){  \includegraphics[width=.8\hsize]{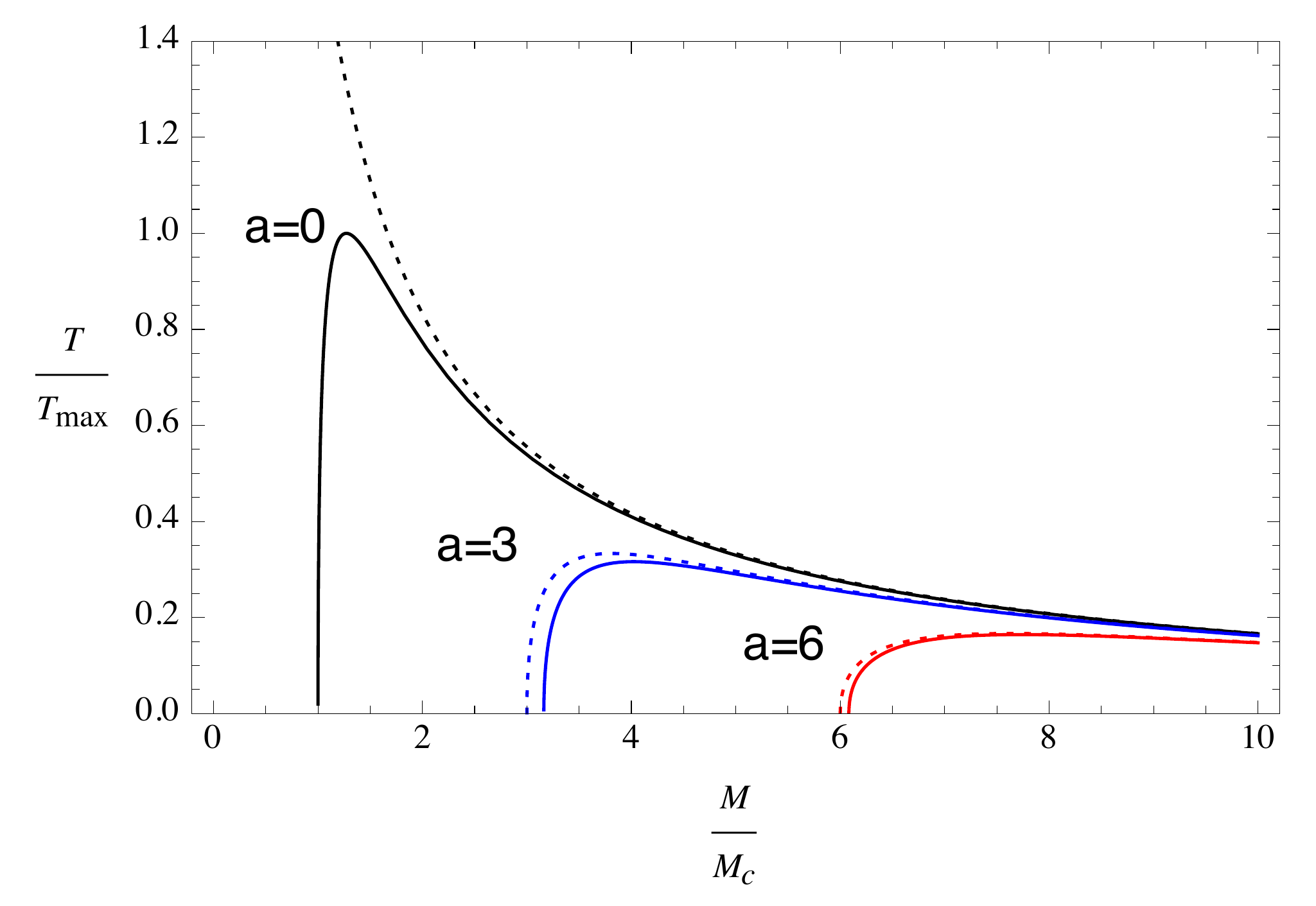}}
\end{picture}
\caption{\label{TempPlot}  {Horizon temperature as a function of the black hole mass, 
comparing classical gravity (dashed lines) with asymptotically safe gravity with $g_*=1$ 
(solid lines) for several angular momenta, with $a$ given in units of $1/M_c$. Temperatures 
are normalised to the maximum temperature of the asymptotically safe Schwarzschild black 
hole (see text).}}
 \vskip-.3cm
\label{pKerr}
\end{center}\end{figure*} 
\subsection{Temperature and specific heat}\label{Temperature}
For given mass of the black hole $M\geq M_c$, the equation of state displays two solutions for the area $A$, and consequently, for the temperature $T$. These can be interpreted as the outer $A\geq A_c$ and inner $A\leq A_c$ horizons of the black hole with mass $M$. The temperature $T$ of the black hole  with $A\geq A_c$ follows from \eq{MAS} or \eq{ASarea} through 
appropriate  differentiation. In Fig.~\ref{TempPlot} we show the temperature \eq{T} of the 
black hole for the outer horizon for various values of the rotation parameter $a = J/M$. In all 
cases, and in contradistinction to the classical Schwarzschild black hole, the  temperature 
falls to zero in the limit $M \to M_c$. This pattern implies the existence of a maximum temperature 
which at $J= 0$ is found to scale as 
\beq
T_{\rm max} \propto \sqrt{g_*}M_P=E_c\,.
\eeq
With \eq{Gk} the proportionality factor reads $(1+\sqrt{5})^{1/2}/(2^{3/2}(2+\sqrt{5})\pi)\approx 0.024$ 
showing that the largest achievable temperature stays well below Planckian energies for all $M$, 
provided that $g_*$ is of order unity. Note that the horizon with $A>A_c$ does not encode UV scaling, because it relates only to modes with $k<E_c$ \eq{kc}. Therefore the UV limit $k\to \infty$ cannot be taken for the branch with $A>A_c$.

The temperature for the inner horizon where $A\leq A_c$ is computed similarly. Here, instead, it relates to modes with $k>E_c$ \eq{kc}, and the scaling limit can be performed. Using \eq{T}, we find that temperature and mass display universal scaling at the fixed point,
\begin{equation}
T\sim k\,,\quad\quad M\sim k
\end{equation}
In contrast, the definition of temperature \eq{T2} which relates to the definition of entropy \eq{deltaSalt} leads to
\begin{equation}\label{T-nonuni}
T\sim \frac{G_N\, k^3}{g_*}\,,\quad\quad M\sim k\,.
\end{equation}
This scaling behaviour is non-universal as it refers to an additional dimensionful quantity, $G_N$, the value of Newton's constant in the IR, other than the RG scale $k$. From an RG perspective, the non-universality is understood from the fact that the temperature coefficient in \eq{T-nonuni} is proportional to $g/\beta_g$, which diverges when approaching the UV fixed point ($g=Gk^2$ and $\beta_g\equiv \partial g/\partial \ln k$). The additional factor $\beta_g/g$ is an RG scheme-dependent non-universal quantity.  This provides additional support to adopt \eq{deltaS} rather than \eq{deltaSalt}.

The specific heat associated to the black hole is defined as 
\beq\label{C}
C =\frac{\partial M}{\partial T}\,.
\eeq
In Fig.~\ref{HeatPlot} we show the specific heat  \eq{C} in comparison with the classical result 
(dashed lines) for different angular momenta. For vanishing angular momenta, the classical 
specific heat is always negative. Once RG effects are taken into account, the specific heat 
changes sign for black hole masses approaching $M_c$. This happens in a regime where the 
thermodynamical approximation is viable, and  thus is a prediction of our theory. It implies a qualitative change in the 
thermodynamics in that the black hole becomes thermodynamically stable. The specific heat vanishes once 
its mass is as low as $M=M_c$ allowing for  a cold black hole remnant. Furthermore, for non-vanishing 
angular momenta, classical black holes show a change in specific heat for sufficiently small black hole 
masses. Including quantum corrections, we note that the sign flip in the specific heat takes place already 
at larger masses. Furthermore, the critical black hole mass is also larger than in the classical case.

  \begin{figure*}[t]
\begin{center}
\unitlength0.001\hsize
\begin{picture}(800,570)
\put(0,0){\includegraphics[width=.8\hsize]{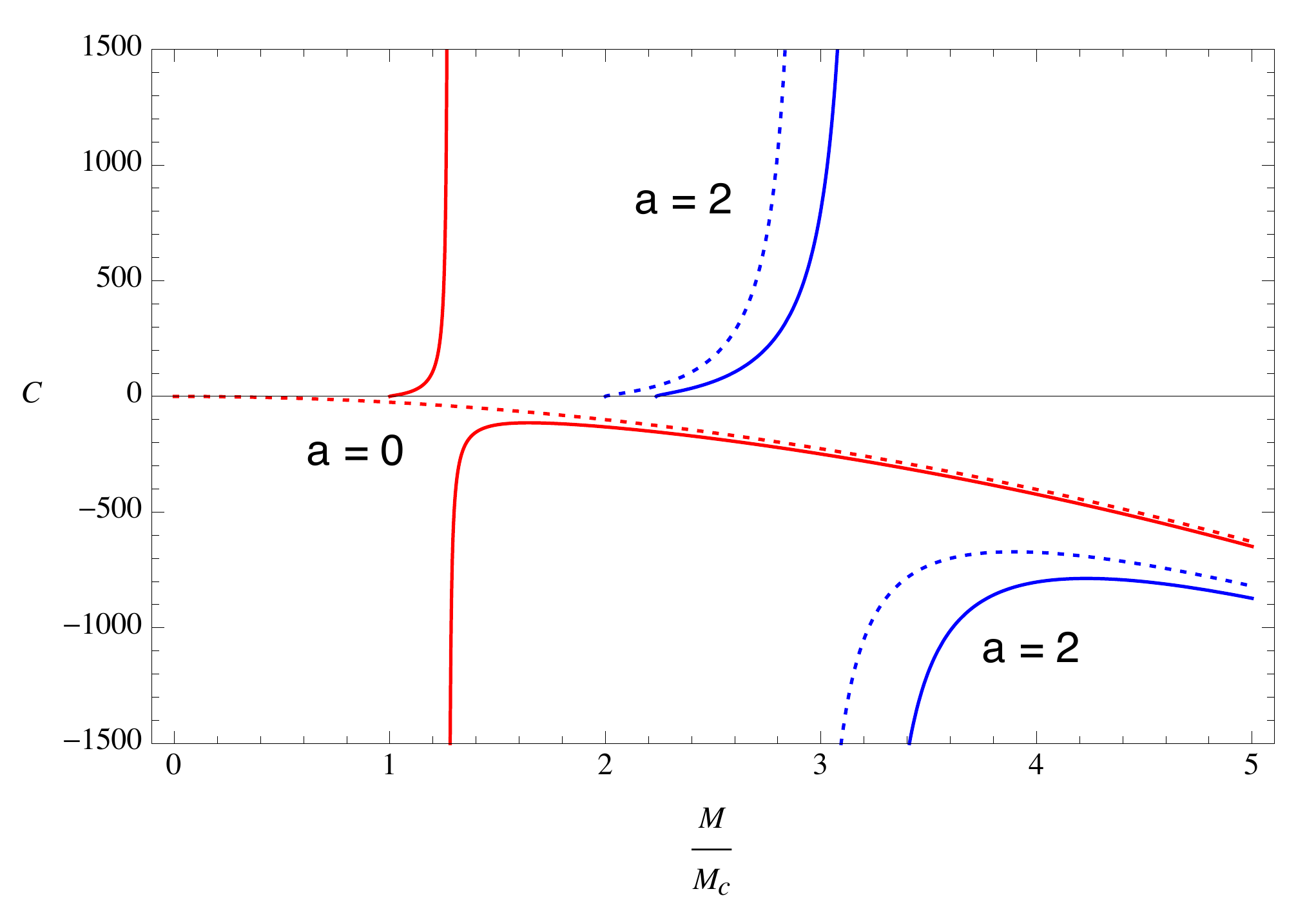}}
\end{picture}
\caption{\label{HeatPlot}  {Specific heat as a function of the black hole mass, comparing 
classical gravity (dashed lines) with asymptotically safe gravity  ($g_*=1$, solid lines) for 
several angular momenta $a$, given in units of $1/M_c$  (see text).}}
 \vskip-.3cm
\label{pKerr}
\end{center}\end{figure*} 

\subsection{Inverse mass expansion}
It is interesting to perform an expansion in powers of $M^2_c/M^2$, which corresponds to an expansion in powers of $\hbar$, see Sec.~\ref{FP}. This is achieved 
by either sending $M_c\to 0$ at fixed black hole mass $M$, 
or by sending the mass to infinity $1/M\to 0$ at fixed $M_c$. We adopt the RG running 
\eq{Gk}. Expanding the state function in units of the classical state function, we find
\beq\label{Aexpansion}
\frac{A}{A_{\rm cl}}=1-\frac12\left(\frac{M_c}{M}\right)^2
-\frac1{16}\left(\frac{M_c}{M}\right)^4
-\frac{1}4\left(\frac{M_{\rm Kerr}}{M}\right)^4\left(\frac{M_c}{M}\right)^2
+{\rm subleading}\,,
\eeq
using \eq{Mc} and \eq{MKerr}. Here we have also introduced  the area of the classical horizon $A_{\rm cl}$, which for 
$J=0$ reads $A_{\rm cl}=4\pi(2\,G_N\,M)^2$. Note that the expansion is a power series 
in $M^2_c/M^2$ times kinematical corrections in powers of $M^2_{\rm Kerr}/M^2$. All subleading terms 
originate from fluctuations and decrease the horizon area relative to 
the classical horizon at the same mass and angular momentum.   
The first two terms are independent of angular momentum. 
The ratio \eq{Aexpansion} interpolates between 
$1$ in the classical limit and $\frac14$ in the limit where the black hole becomes critical $M\to M_c$.  
Similarly, for the temperature we find
\beq\label{Texpansion}
\frac{T}{T_{\rm cl}}=1-\frac14\left(\frac{M_c}{M}\right)^2
-\frac5{16}\left(\frac{M_c}{M}\right)^4
-\frac{5}{16}\left(\frac{M_{\rm Kerr}}{M}\right)^4\left(\frac{M_c}{M}\right)^2
+{\rm subleading}\,,
\eeq
showing that quantum corrections decrease the temperature in comparison to the 
classical one. Here, $T_{\rm cl}$ denotes the classical temperature of the black hole 
which reads $T_{\rm cl}=M_P^2/(8\pi\,M)$ for $J=0$. The  corrections to \eq{Aexpansion} 
and \eq{Texpansion} are algebraic, which is a consequence of the power-law running 
of Newton's coupling under the RG flow \eq{Gk}.

\subsection{Conformal scaling}\label{CS}

We now turn to the conformal scaling laws of black holes within asymptotically safe quantum 
gravity 
in the vicinity of an ultraviolet fixed point.  
Under the assumption that  the underlying partition function at high energies is 
dominated by semi-classical black holes, it has been suggested by O.~Aharony 
and T.~Banks \cite{Aharony:1999uq} and by A.~Shomer \cite{Shomer:2007kx} 
that a quantum theory of metric gravity may not exist as a local quantum field 
theory. Here, we re-evaluate this line of argument in the  light of the asymptotic safety 
conjecture. 

For want of generality we consider the case for black holes in general 
dimension $d$, and take $J=0$ for simplicity. We recall that for a conformal field 
theory (CFT),  the entropy and energy scale as
\beq \label{CFT}
S \sim (R T)^{d-1} , \,\,\,\,\,\,\, E \sim R^{d-1} T^d
\eeq
where $R$ is the radius of spacetime under consideration, and $T$ is the temperature.  
It is important when dealing with black holes to note that the black hole radius $R$ 
depends on the energy $E=M$ of the black hole. Therefore we should consider a 
relation between the entropy and energy densities of the form
\beq \label{scaling}
\frac{S}{R^{d-1}} \sim \left( \frac{E}{R^{d-1}} \right)^{\nu}\,.
\eeq
For a conformal field theory, the scaling behaviour  \eq{CFT} dictates \eq{scaling} with
\beq\label{conformal}
\nu_{{}_{\rm CFT}} = \frac{d-1}{d}
\eeq
and $T^{d-1}\sim {S}/{R^{d-1}} $. The scaling relation \eq{conformal} is different from 
the one put forward by A.~Shomer \cite{Shomer:2007kx}, according to which 
entropy scales with energy as $S \sim E^{\frac{d-1}{d}}$. The latter would only be true 
if the radius was independent of the mass and entropy. This is not the case for black 
holes such as those considered here. For a semi-classical black hole we have 
that  $A \sim R^{d-2}$, $E \sim G_N^{-1} R^{d-3}$ and $S \sim R^{d-2} G_N^{-1}$,  
where $R$ is the Schwarzschild radius, leading to the scaling relation \eq{scaling} with index
\beq\label{semi}
\nu_{{}_{\rm BH}}=\frac12
\eeq 
for any dimension. Not surprisingly, \eq{semi} shows that (semi-)classical  black holes 
do not behave as conformal field theories. This  also follows from the fact that 
the Schwarzschild solution depends on the dimensionful quantity $G_N$, implying 
that  the physics cannot be scale invariant. On the other hand, extrapolating down to 
two dimensions where  $G_N$ is dimensionless, we find that the semi-classical estimate \eq{semi} is formally in agreement with conformal scaling \eq{conformal}.

   \begin{figure*}[t]
\begin{center}
\unitlength0.001\hsize
\begin{picture}(750,500)
\put(0,0){  \includegraphics[width=.7\hsize]{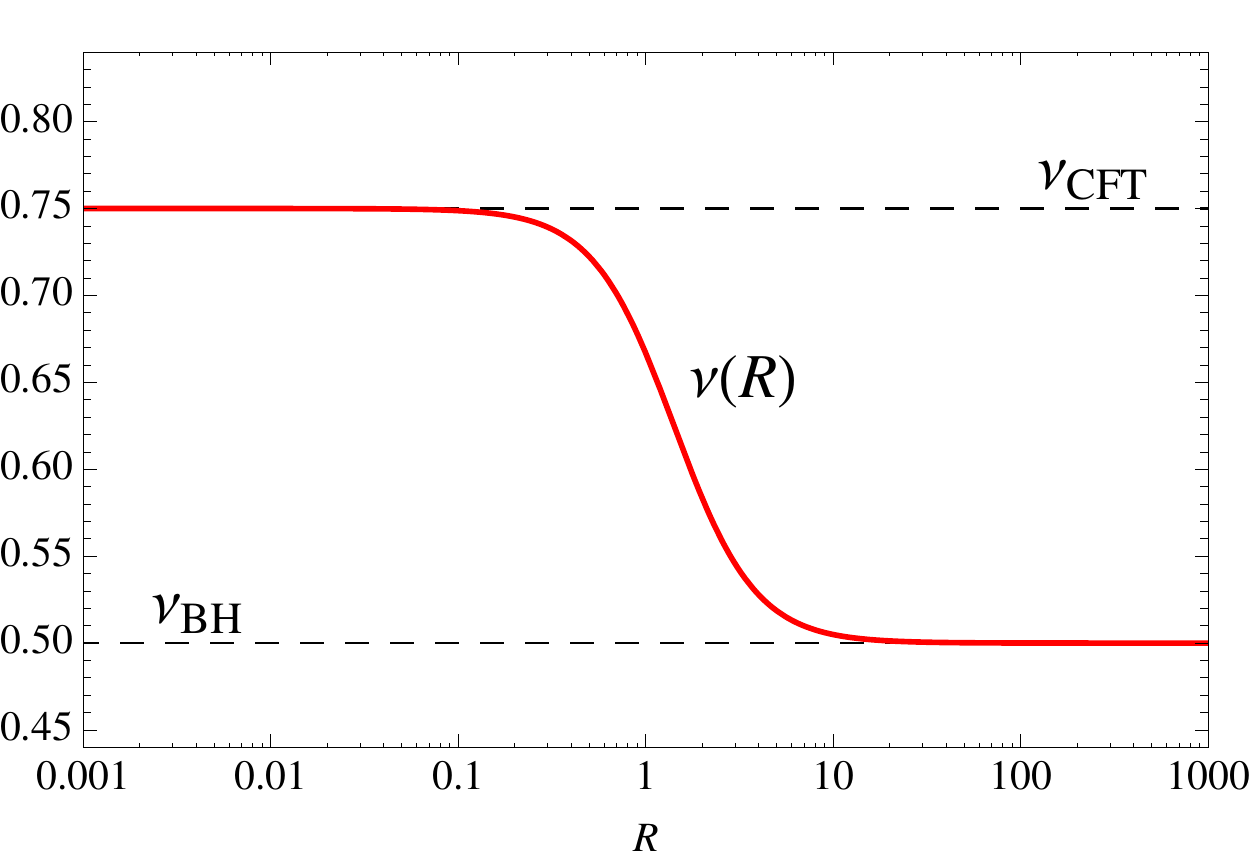}}
\end{picture}
\caption{\label{ScalingPlot}  {Scaling index for an asymptotically safe  Schwarzschild black hole 
in four dimensions interpolating between the classical value $\nu_{{}_{\rm BH}}$ for large horizon radii and the conformal limit $\nu_{{}_{\rm CFT}}$ for small radii. The radius $R$ is given in units of $R_c$.}}
 \vskip-.3cm
\label{pScaling}
\end{center}\end{figure*}

We now turn to the scaling of asymptotically safe black holes. The central observation is that  
the horizon area always scales  according to $A\sim k^{2-d}$, and hence $R\sim k^{-1}$. 
For $k\ll E_c$, energy, entropy  and temperature scale exactly the same way as in the 
classical case, leading to \eq{semi}. For $k\gg E_c$, fixed point scaling takes over and we  
find that entropy becomes a constant while both mass and temperature scale linearly with 
energy $M \sim k$ and $T \sim k$ in this limit, leading to
\beq\label{conformalBH}
T\sim R^{-1}\,,\quad E = M \sim R^{-1}  \quad  {\rm and } \quad S = {\rm const.}
\eeq
for asymptotically safe black holes in any dimension. The scaling \eq{conformalBH} is 
evidently conformal, obeying \eq{scaling} with $\nu$ given by \eq{conformal}. The appearance 
of conformal scaling can also be understood by noting that the gravitational ultraviolet fixed point  removes the infrared scale $G_N$ 
from the set-up. Consequently, in the absence of any other scales, the system must  fall back 
onto \eq{scaling} for any dimension.  In Fig.~\ref{pScaling}, we have computed the index
\beq\label{nu}
\nu = \left( d-1 -\frac{\partial \ln S}{\partial \ln R}\right) 
\left(d-1-\frac{\partial \ln E}{\partial \ln R}\right)^{-1}  \,.
\eeq
in four dimensions along the RG trajectory \eq{Gk}, with $R_c$ denoting the scale corresponding 
to $A=A_c$ and $k=E_c$.  With decreasing $R$, the index shows a smooth cross-over from 
classical behaviour for large $R$ to conformal scaling for small $R$.  Interestingly, this result is also consistent with a holographic view on the RG, see \cite{Litim:2011qf,Litim:2012zz}. We conclude that the UV fixed point scaling 
of asymptotically safe quantum gravity  is encoded in the 
Cauchy horizon of its black hole solutions. 

\section{Entropy and black hole space-times} \label{secIV}
In this section we provide explicit space-time metrics which carry the 
thermodynamics derived in the previous 
sections.. 
We also relate our findings with earlier work based on 
RG improved metrics, discuss the entropy and its quantum corrections,  
and a compare results with other approaches to quantum gravity.

\subsection{Metrics}
The construction of the previous sections makes no reference to an explicit underlying space-time 
metric. For some applications, it will be useful to have explicit RG improved metrics available 
which carry the thermodynamics derived above. In fact, it is possible to provide such metrics 
for any choice of coordinates. As an example, we consider the Kerr metric for an uncharged 
black hole $(q=0)$ in the familiar Boyer-Lindquist coordinates,
\begin{eqnarray} \label{kerrmetric}
ds^2 & = & -\left(1- \frac{2G\,M\, r}{\rho^2(r)} \right) dt^2 
- \frac{G\,M\,r }{\rho^2}\,a\,\sin^2\theta\, dt\, d\phi +
\frac{\rho^2(r)}{\Delta(r)} dr^2 + \rho^2(r) d \theta^2 \nonumber\\
& &+  \frac{\sin^2\theta}{\rho^2(r)} 
\left[ (r^2 + a^2)^2 - a^2 \Delta(r) \,\sin^2\theta  \right] d\phi^2 
\end{eqnarray}
where $a= \frac{J}{M}$ denotes the angular momentum in units of the mass, and
\begin{eqnarray}
 \Delta(r) &=&  r^2 -2G\,M r +a^2\\
 \rho^2(r) &=& r^2 + a^2 \cos^2 \theta \,.
\end{eqnarray}
The horizons radii are found from solving $\Delta(r_\pm)= 0$ with $r_+$ and $r_-$ 
the well-known outer and inner horizon, respectively, and the horizon area is then given 
by $A=4\pi^2(r_\pm^2+a^2)$. In the classical theory $G$ is a constant, given by 
Newton's coupling. In the spirit of a renormalization group improvement, we now wish 
to take the RG running of couplings into account, replacing
\beq\label{Gr}
G\to G(r,\cdots)
\eeq  
where the new coupling $G(r,\cdots)$ depends on the coordinates and parameters of 
the space-time metric such as the radial distance $r$. We expect, by continuity, that 
changes in  the numerical value of $G$ in \eq{kerrmetric} along some RG trajectory 
account for the leading corrections to the effective space-time geometry. The  RG equations 
provide us with the scale-dependence of couplings $G\to G(k)$,  and the 
coordinate-dependence of couplings is then imported  by means of a scale identification
\beq\label{identify}
 k=k(r,\theta;a,M)\,.
 \eeq 
Thus, the scale identification \eq{identify} is a central ingredient because it affects the physical properties of 
RG improved black hole metrics. In a multiscale problem, deriving an suitable definition for $k$ in terms of the physical mass parameters is a highly non-trivial task \cite{Bonanno:2000ep, Falls:2010he,Anber:2011ut}. We claim, however, that the scale identification
\beq \label{kr}
k^2 \sim \frac{1}{r^2 + a^2}
\eeq
is distinguished as it leads to an RG improved black hole space-time with identical thermodynamical 
relations as those derived in Sec.~\ref{secII} in a metric-independent manner. The identification implies that  one 
recovers \eq{k} and hence $G(r_\pm) = G_{\rm opt}(A)$ on the horizons $r \to r_\pm$.

 \subsection{Thermodynamics}\label{thermo}

We establish the thermodynamical equivalence between RG improved black hole metrics 
with \eq{Gr} and  \eq{kr} and the RG thermodynamics derived in Sec.~\ref{secII}. The equivalence 
is such that the relation between $M$, $J$, $q$ and $A$ given by \eq{Mimp}  is satisfied, and 
that the temperature \eq{T} corresponds exactly to the surface gravity of the RG improved black 
hole metric, i.e. $T= \frac{\kappa}{2 \pi}$. Our reasoning is independent of the specific  RG 
scale dependence of couplings. We consider the example of the Kerr-Newman black hole, 
and begin by replacing the couplings through running couplings using \eq{kr}. We denote 
them as $G(r)$ and $e^2(r)$, although they also depend on $a$. The RG improved equations 
for the Kerr-Newman black hole follow from  the Kerr metric \eq{kerrmetric}, substituting
$2G M r$ by $2G(r) M r - G(r) e^2(r) q^2$. The horizon condition at radial coordinate $r= r_+$ 
is now given by $\Delta(r_+)=0$ where
\beq\label{Delta}
\Delta(r) =  r^2 + a^2 - 2G(r) M r  + G(r) e^2(r)q^2 \,.
\eeq
The area of the black hole event horizon reads $A= 4 \pi (r_+^2 +a^2)$  in terms of the rotation 
parameter $a$ and $r_+$. From $\Delta=0$ we have the relation
 \beq \label{rplusKN}
 r_+ = \frac{A+ 4 \pi e^2(r_+) q^2 G(r_+)}{8 \pi M G(r_+)} \, .
 \eeq
One then finds a state function which relates mass with angular momentum, charge and the area
\beq\label{mass}
M^2 = \frac{4 \pi}{A} \left[ \left(  \frac{A+ 4 \pi e^2(r_+) q^2 G(r_+)}{8 \pi  G(r_+)}\right)^2 + J^2 \right]\, .
\eeq
Upon the use of \eq{kr}, and hence $G(r_+)= G_{\rm opt}(A)$ and $e^2(r_+) = e^2_{\rm opt}(A)$,  
we find that the state function \eq{mass} agrees with \eq{Mimp}. Since the functional dependence 
of $M(A,J,q)$, as given by  \eq{Mimp}, on $J$ and $q$ is the same as for a classical black hole we 
find that the potentials $\Omega$ and $\Phi$ obtained by taking derivatives of $M$ equally retain 
their classical form,  the only difference being that $e^2$ and $G_N$ are replaced by the running 
couplings, and the classical horizon radius replaced by $r_+$. Expressed in terms of $r_+$ and 
$a$, the potentials
\begin{eqnarray}
\Omega&=& \frac{\partial M}{\partial J} = \frac{a}{r_+^2+a^2}\\
\Phi &=& \frac{1}{e(r_+)}  \frac{\partial M}{\partial q} = e(r_+) q \frac{r_+}{r_+^2+a^2} 
\end{eqnarray}
agree with the expressions obtained from the metric and the RG improved electric potential.
 Finally, we turn to the black hole temperature. In the metric formulation it is given by the surface gravity 
 on the black hole horizon $T=\frac{\kappa}{2 \pi}\equiv  \frac{1}{4 \pi} \frac{\Delta'(r_+) }{r_+^2 +a^2}$. 
 Using \eq{Delta}, we find that
 \bea
 T 
 &= & \frac{1}{4 \pi r_+} \left[ \frac{r_+^2-a^2}{r_+^2 + a^2}  -  \frac{r_+}{ G(r_+)}  G'(r_+)\right.  
 \label{Tmetric}
 \left. \frac{e^2(r_+)q^2 G(r_+)}{r_+^2 + a^2} \left( 1- \frac{r_+}{ e^2(r_+)} e^2\, '(r_+)\right) \right] \, ,
 \eea
where primes denote derivatives with respect to the argument.
We have to show that this expression is equivalent to the temperature defined in  \eq{T},  
$T= 4 G(A){\partial M}/{\partial A}\,.$
Using the mass function  \eq{Mimp} as well as \eq{rplusKN}, we find explicitly
 \bea\label{Tnonmetric}
& &T
=\frac{1}{ 4 \pi r_+} \left[  \frac{r_+^2-a^2}{r_+^2 + a^2}  
 -   \frac{2 r_+^2}{r_+^2 + a^2}\frac{\partial \ln G_{\rm opt}}{\partial \ln A}
 - \frac{G_{\rm opt}e^2_{\rm opt}q^2}{r_+^2+ a^2}  
\left(1-\frac{2 r_+^2}{r_+^2 + a^2} \frac{\partial \ln e^2_{\rm opt}}{\partial\ln A}   \right)\right]
\label{dMdA} 
\eea
Clearly, \eq{Tmetric} and \eq{Tnonmetric} agree in the absence of RG corrections. In the presence 
of non-trivially running couplings, the terms involving derivatives of couplings have to agree as well. 
Here,  in consequence of the scale identification \eq{k} and \eq{kr}, we have that 
\beq\label{relation}
r\,\partial_r\large|_{r=r_+} = \frac{2 r^2_+}{r^2_+ + a^2}\, A\, \partial_A
\eeq
when applied on the running couplings.  Using \eq{relation} we therefore conclude that  
\eq{Tmetric} and \eq{Tnonmetric} are identical, term by term, as claimed.

It is worth pointing out that the equivalence of the RG-improved state functions and temperature as induced by RG-improved metrics \eq{kerrmetric} with \eq{Gr}, \eq{kr} with the metric-independent derivation of thermodynamical relations given in the preceeding sections is non-trivial. This includes a regime of conformal scaling, provided the RG flow displays a UV fixed point.  We also stress that that the equivalence of state functions and temperature functions \eq{Tmetric} and \eq{Tnonmetric} strengthens our choice \eq{deltaS}, \eq{T} over \eq{deltaSalt}, \eq{T2}. In fact, for the latter choice the RG-improved thermodynamics and the RG-improved metrics would have led to different results.

Furthermore, one cannot expect that \eq{kerrmetric} with \eq{Gr} and a generic matching 
necessarily leads to a picture compatible with the relations of black hole thermodynamics. In the literature, physically motivated 
 matching conditions have been explored including $k\sim 1/r$ 
 \cite{Bonanno:2000ep, Falls:2010he, Cai:2010zh, Burschil:2010ys, Reuter:2011uq}, or 
 $k\sim r_{\rm cl}^{\gamma-1}/r^\gamma$ for some model parameter $\gamma$ \cite{Falls:2010he}, 
 and matchings $k\sim 1/D$ \cite{Bonanno:2000ep, Bonanno:2006eu,Falls:2010he} 
 where $D(r ,\theta)$ denotes the proper distance of the classical space-time.  The RG 
 improved metrics for all matchings studied thus far consistently predict the existence of a smallest Planck-size black 
 hole. However, for rotating black holes, none of these obey \eq{relation} and the related  metrics fail 
 to reproduce \eq{state} or equality of the temperatures  \eq{Tmetric} and \eq{T}. Moreover, in these 
 cases one cannot define an entropy function without giving up the relation $T=\frac{\kappa}{2 \pi}$ 
 since the $1$-form $\delta Q/T$ is neither exact nor an integrating factor can be found \cite{Reuter:2011uq}.  
 In turn, the scale identification \eq{kr} resolves these matters. For Schwarzschild black holes this aspect is 
 hidden as the relation \eq{relation} becomes less restrictive. Then matchings of the form $k\sim 1/r$ lead 
 to a consistent thermodynamics, and the $1$-form $\delta Q/T$ is trivially exact.

\subsection{Entropy}\label{Entropy}

Next we turn to  the entropy and its quantum corrections in the light of the RG, and compare the thermodynamical and statistical entropy with  Clausius' definition, and with results from  the  literature based on other approaches. 

In the absence of quantum gravity effects, the classical Bekenstein-Hawking result \eq{entropy} states that the black hole's entropy is larger, at fixed area, the smaller the classical coupling $G_N$ (and vice versa). Under the renormalisation group flow, the entropy is modified, \eq{Sk}. Within the asymptotic safety scenario, the key quantum gravitational effect is that the running Newton coupling decreases (increases) with increasing (decreasing) RG momentum  scale, provided the fixed point $g_*$ is positive (negative). Therefore we  expect that the quantum corrections to the entropy have the same sign as $g_*$.

More specifically, inserting the  non-perturbative RG running \eq{Gk} into \eq{Simp},  
and also using \eq{k}, the thermodynamical  entropy reads
\beq\label{SimpA}
S=\frac{A}{4 G_N} + \frac{\pi}{g_*} \,.
\eeq
The result applies equally for non-rotating, rotating or charged black holes. The constant term is a fingerprint of the underlying fixed point. There are no subleading terms in inverse powers of the area but they could arise from more sophisticated approximations for the RG flow \cite{Litim:2003vp}. 
Unlike the Bekenstein-Hawking entropy \eq{entropy}, the quantum-corrected expression \eq{SimpA} remains strictly positive  even for vanishing area due to the non-perturbative RG running of  Newton's coupling and $g_*>0$. The constant term has a quantum origin and contains an additional power of $\hbar$ compared to the leading term.
The entropy is dominated by the first term for large masses and horizon areas, and the quantum corrections are parametrically suppressed both as $\propto G_N/A$ and as $\propto \hbar$. In the limit where the mass approaches the critical mass $M_c$, the entropy approaches the 
value  
\begin{equation}
\label{Sc}
S_c=2\pi/g_*\,.
\end{equation}
For $g_*$  of the order one, as found 
in explicit RG studies, the entropy $S_c$ of a critical  black hole is of the order of a few. Hence, the numerical value of the gravitational fixed point determines  the effective number of degrees of freedom of a quantum black hole with mass $M_c$. In the asymptotic limit of vanishing area, the theory becomes conformal and the entropy a constant of the order of a few, see Sec.~\ref{CS}. This limit can also be achieved for parametrically small $g_*\to 0$  at fixed area, corresponding to a regime of quantum dominance where $1/\hbar\to 0$.
We also note the absence of logarithmic corrections 
to the expression given in \eq{SimpA}. This is so because the entropy expression \eq{SimpA} ultimately arises from summing over all horizon 
areas  \cite{Carlip:1993sa}. The additional area-dependence entering through the RG running of the 
Newton's coupling, responsible for the constant term, is not generating a logarithm, although it could have done so, provided that the gravitational RG running \eq{dG} explicitly receives logarithmic dependences on the RG scale. 

If, on the other hand, we use
Clausius' phenomenological definition for the entropy, it follows from \eq{T} (which holds true for the RG improved metric) that
\beq \label{metricentropy}
S=\int dS = \int \frac{dM}{T}  = \int \frac{dA}{4G(A)}\,.
\eeq
The total derivative of
\eq{metricentropy} 
leads to
\eq{deltaS}, and the entropy arises as a weighted sum over all areas, where the running of the inverse Newton coupling with area 
serves as the weighting factor. Clausius' rule assumes that we can straightforwardly compare the entropy 
of two black hole solutions with thermodynamics defined at different coarse-graining scales $k$ 
evaluated at the horizon.  
Performing the integral in \eq{metricentropy}, 
and using the asymptotically safe RG running for $G_k$ given by \eq{Gk} together with \eq{k},  we find
a logarithmic correction to the entropy, which we express as an entropy difference
\beq\label{Smetric}
S(A)-S(A_c) = \frac{A-A_c}{4 G_N} + 
\frac{ \pi}{g_*} \,\ln \left(\frac{A}{A_c}\right)\,,
\eeq
where $A_c$ serves as a reference point. Clausius'  entropy \eq{Smetric} is  quite general 
in that it applies universally for rotating and charged black holes, despite of being only a function 
of the area $A$.
The result falls back onto \eq{entropy} in the limit of large areas where quantum corrections are subleading as $(\ln A)/A$. The main point  is that \eq{Smetric} displays logarithmic quantum corrections linear in $\hbar$ in addition to the terms found in \eq{SimpA}. The sign of the proportionality factor is fixed by the sign of $g_*$ for $A\ge A_c$. Note that for $g_*<0$, a cancellation may occur between the two terms in \eq{Smetric} at $A\neq A_c$. Note also that the expression still requires input for $S(A_c)$ which is not determined by \eq{metricentropy} alone. If we require that \eq{Sc} holds true for both \eq{SimpA} and \eq{metricentropy} at $A=A_c$, we find
\begin{equation}\label{Scc}
S(A)= \frac{A}{4 G_N} + \frac{\pi}{g_*}\left(1+ \ln 
\frac{A}{A_c}
\right)\,.
\end{equation}
We conclude that the entropy 
\eq{Smetric}, \eq{Scc} derived phenomenologically using Clausius' rule
differs by a logarithm from the expression \eq{SimpA}  to which we were lead via thermodynamical considerations.

In order to gain further insights into an appropriate definition of the entropy, we also compute the statistical entropy 
of the RG improved metric obtained from the functional integral.  This can be done using the ``off shell'' 
conical singularity method by S.~Solodukhin \cite{Solodukhin:1996vx} for the RG improved 
Schwarzschild black hole $J=q=0$.
To that end we approximate the 
Euclidean action by \eq{action} plus the Gibbons-Hawking surface term, with $k= k_{\rm opt}(A)$ 
according to \eq{k}. From this one obtains the free energy $F \equiv T\, \Gamma^E$, where 
$\Gamma^E$ denotes the Euclidean effective action at the scale \eq{k}. Inserting the RG improved metric with \eq{kr} into the 
action we find that the free energy is given in terms of mass, temperature and entropy as
\beq\label{F}
F=  M  - S\, T
\eeq
for all RG scales. Here, the mass $M$ is given explicitly by the mass function \eq{Mimp} (with $J=q=0$), and the 
entropy is given by \eq{Simp}. From the validity of \eq{F} we conclude that the statistical definition of entropy for the RG improved black hole space-time agrees with the thermodynamical definition of entropy, \eq{Simp}. This result also shows that the statistical defintion of entropy differs by a logarithmic term from the one obtained by applying Clausius' rule. 

\subsection{Comparison}

Finally, we compare our findings with selected results from the literature, see \cite{Wald:1999vt} for an overview. There is a vast body of work dealing with quantum corrections to black hole entropy,  including 
applications of the  conical singularity method in Euclidean space-times and relations to the conformal 
anomaly \cite{Fursaev:1994te, Frolov:1996hd, Mann:1997hm,Solodukhin:2011gn,Sen:2012dw}, the use of Cardy's formula 
\cite{Carlip:2000nv}, or studies of backreaction effects \cite{Lousto:1988sp}. Logarithmic corrections to 
the entropy have also been found in the loop quantum gravity approach \cite{Kaul:2000kf},  perturbative 
quantum gravity and in string theory  
\cite{Susskind:1994sm,Fursaev:1994ea,Kabat:1995eq,Larsen:1995ax,Sen:2012cj}, or based on 
phenomenologically motivated expansions of classical space-times in powers of $\hbar$ \cite{Akbar:2010nq}. 
No general agreement has yet been achieved neither for the coefficient nor the sign of the logarithmic term, which has 
even been conjectured to be absent altogether \cite{Medved:2004eh}.  

In spite of this, some structural insights relate with our findings and are worth being highlighted. Firstly, for Clausius' 
entropy defined in \eq{Smetric} and \eq{Scc}, the logarithmic correction can in principle have either sign depending on 
whether the RG running \eq{Gk} is dominated by gravitational fluctuations or external matter fields. 
A positive (negative) sign correlates with the existence (absence) of a maximum black hole 
temperature. This very same link was noticed earlier in different settings, e.g.~in \cite{Lousto:1988sp} by analysing back-reaction 
effects, and in \cite{Akbar:2010nq} by relating the logarithmic coefficient to the leading quantum correction for 
the surface gravity. 
Secondly,    statistical 
considerations  have been used in \cite{Hod:2004cd}  to  argue that  the coefficient of the logarithmic term 
should be positive or vanishing. This constraint is in accord with our findings 
as long as gravity is anti-screening as implied by asymptotic safety. Finally, the  sign pattern found here is in agreement with results from perturbative 
gravity \cite{Susskind:1994sm} and the perturbative RG \cite{Fursaev:1994ea,Kabat:1995eq,Larsen:1995ax}. Within loop quantum gravity \cite{Ashtekar:1997yu}, the logarithmic correction has apparently the opposite sign, a result which has been interpreted as a hint towards an underlying non-perturbative fixed point  \cite{Kaul:2000kf}. 

The Wilsonian perspective developed  here  can also be applied for other UV completions of gravity. As such it may offer a common framework to study  similarities and differences between approaches to quantum gravity.

\section{Discussion} \label{secV}

We have put forward an approach to understand the thermodynamics of black holes from a renormalisation group perspective,
allowing for a continuous interpolation between black holes with largest and smallest mass. The main new ingredient is the scale-dependence of couplings such as Newton's constant $G\to G(k)$ or the fine-structure constant $e^2\to e^2(k)$ arising from the underlying effective action  \eq{action}.
We then find that thermodynamics imposes a relation between 
the RG scale and  the horizon area of the black hole \eq{k}, 
which acts as a diffeomorphism-invariant cutoff for the effective action. The running couplings turn into functions of the black hole horizon area $G\to G(A)$ and $e^2\to e^2(A)$
leading to modifications for the black hole equation of state.  This is consistent with the view that the degrees of freedom with a characteristic wavelength set by the horizon scale, or smaller, are responsible for the thermodynamical nature of black holes.  In this light, our set-up offers
an interpretation for the emergence of black hole thermodynamics 
by treating the background gravitational field  as a coarse-grained field which arises from modes
with wavelengths 
bound by the horizon scale.

On a practical level, our set-up translates the RG-induced modifications into corrections for temperature, specific heat, and entropy without making assumptions about the actual RG running of couplings.
If quantum gravity is (anti-)screening, we find that quantum corrections (decrease) increase the black hole temperature at fixed mass, charge, and angular momentum, as well as the entropy at fixed horizon area.  It is conceivable that the broad picture, which relates the sign of the gravitational $\beta$-function with the sign of quantum corrections to temperature and entropy,  persists in approximations beyond those adopted here. 
We also stress that our equations are completely general
for stationary black hole solutions to Einstein-Maxwell gravity in four dimensions. 
For known classical black hole solutions, these are straightforwardly generalised to equations of state for black holes in dimensions different from four.

Provided, additionally, that gravity becomes asymptotically safe,
the equation of state implies that the temperature is always smaller than the classical temperature for the same mass, angular momentum and charge. Furthermore, the temperature displays a maximum,
and the specific heat of small  black holes becomes positive, see Figs.~\ref{TempPlot} and~\ref{HeatPlot}. The new equation of state also predicts the existence of a lightest black hole for a causal horizon to exist.
Interestingly, the weakening of gravity increases the entropy in comparison to the semi-classical result \eq{entropy} for the same area, thereby enhancing the domain of validity for a thermal description \eq{validity} towards smaller black hole masses. On the other hand, the thermodynamical picture may be called into question for
near-critical black holes, where specific heat and temperature become small and the entropy parameterically of order unity. This regime would benefit from complementary studies. 

We  also showed that conformal scaling, a fingerprint of an RG fixed point, is  encoded in the equation of state in the limit of vanishing horizon area. This regime can be viewed as the Cauchy horizon of the corresponding black hole space-time metrics. The result strengthens the view that asymptotic safety qualifies as a fundamental quantum theory for gravity.  Furthermore, we have  provided explicit space-time metrics
which carry the same equation of state for all
mass, charge or angular momentum. These findings close a gap in the study of RG-improved black hole metrics, showing that these can accommodate thermodynamical relations even in the charged and rotating case as long as the choice of RG scale is informed by the horizon area of the black hole.

Another interesting question relates to the quantum corrections for the entropy.
For asymptotically safe gravity, we find that the entropy at fixed horizon area increases due to quantum corrections, leading to an entropy of the order of a few in the limit where the black hole becomes critical.  The thermodynamical definition for entropy agrees with the statistical definition of entropy \eq{F} for all RG scales and black hole masses, which serves as a consistency check. On the other hand, deriving the entropy using Clausius' rule leads to an additional term logarithmic  in the area. Qualitatively, the difference arises from identifying the RG scale with the area before, or after, exploiting the effective equations of motion. 
 It will be interesting to relate this difference to other definitions for the entropy, including entanglement entropy or  Wald's entropy \cite{Wald:1993zr}.

\acknowledgements
Results reported in this paper where presented by one of the authors (DL) at the workshop {\it RG approach from cold atoms to the hot QGP}, Yukawa Institute, Kyoto (Aug 2011), at the ESF exploratory workshop {\it Gravity as thermodynamics: towards the microscopic origin of geometry}, Trieste (Sep 2011), at the workshop {\it Quantum gravity, from UV to IR}, CERN (Sep 2011), at the 499th WE-Heraeus workshop {\it Exploring quantum space time}, Bad Honnef (Mar 2012), at the symposium {\it Physics at all scales}, Heidelberg  (Apr 2012), at the workshop {\it Non-perturbative aspects in field theory}, Kings College, London (Nov 2012); and by one of us (KF)  at the University of Graz (May 2012) and at the Technical University Vienna (May 2012). We thank the organisers  for their invitations and hospitality, and the participants for stimulating discussions. This work is supported by the Science and Technology Facilities Council (STFC) [grant number ST/J000477/1].

\bibliography{myrefs}

\end{document}